\documentclass[a4paper, 11pt]{article}
\usepackage{amsmath}
\usepackage{authblk}
\usepackage[margin=2cm]{geometry}
\usepackage{graphicx}
\usepackage{hyperref}
\usepackage[font=small]{caption}

\title{Axonemal regulation by curvature explains sperm flagellar waveform modulation}
\author{M.T.~Gallagher\(^{abc\dagger}\), J.C.~Kirkman-Brown\(^{c}\), D.J.~Smith\(^{bc}\)}
\affil{\footnotesize{\(^{a}\)Centre for Systems Modelling and Quantitative Biomedicine, University of Birmingham\\
        \(^{b}\)School of Mathematics, University of Birmingham\\
        \(^{c}\)Centre for Human Reproductive Science, University of Birmingham and Birmingham Women's and Children's NHS Foundation Trust\\
        \(^{\dagger}\)m.t.gallagher@bham.ac.uk}}
\date{}

\begin{document}
\maketitle
\abstract{
  Flagellar motility is critical to natural and many forms of assisted reproduction.
  Rhythmic beating and wave propagation by the flagellum propels sperm through fluid and enables modulation  between penetrative progressive motion, activated side-to-side yaw and hyperactivated motility associated with detachment from epithelial binding.
  These motility changes occur in response to the properties of the surrounding fluid environment, biochemical activation state, and physiological ligands, however a parsimonious mechanistic explanation of flagellar beat generation that can explain motility modulation is lacking.
  In this paper we present the Axonemal Regulation of Curvature, Hysteretic model (ARCH), a curvature control-type theory based on switching of active moment by local curvature, embedded within a geometrically nonlinear elastic model of the flagellum exhibiting planar flagellar beats, together with nonlocal viscous fluid dynamics.
  The biophysical system is parameterised completely by four dimensionless parameter groupings.
  The effect of parameter variation is explored through computational simulation, revealing beat patterns that are qualitatively representative of penetrative (straight progressive), activated (highly yawing) and hyperactivated (non-progressive) modes.
  Anaylsis of the flagellar limit cycles and associated swimming velocity reveals a cusp catastrophe between progressive and non-progressive modes, and hysteresis in the response to changes in critical curvature parameter.
  Quantitative comparison to experimental data on human sperm exhibiting typical penetrative, activated and hyperactivated beats shows a good fit to the time-average absolute curvature profile along the flagellum, providing evidence that the model is capable of providing a framework for quantitative interpretation of imaging data.}

\section{Introduction}

The migration of spermatozoa to the egg, driven by the active movement of a beating flagellum, is an essential process in the reproductive biology of humans, and many other species. 
Internal fertilisation typically requires sperm to respond to and traverse a range of fluid environments, most notably regions of very high viscosity (and elasticity) such as cervical mucus and the cumulus oophorus surrounding the oocyte. Motility is induced by oscillatory beating and wave propagation of the eukaryotic flagellum, driven by the internal active axoneme structure. 
Despite nearly 70 years elapsing since the discovery of the axoneme, a mechanistic understanding of its control mechanism remains unresolved. 
In particular, modulation of the waveform between penetrative, activated and hyperactivated archetypes is yet to be explained by a parsimonious model. In this paper we reformulate the long-studied concept of curvature control, embedded within a four-parameter, time domain, geometrically nonlinear and hydrodynamically nonlocal mathematical model of the active flagellum.
All three planar archetypes will be found to emerge from this simple model, providing evidence in favour of the curvature control hypothesis and a quantitative framework to investigate motility modulation, including mechanistic parameters for fitting and characterisation of flagellar imaging data.

The question as to how the flagellum is internally regulated has been discussed since at least as early as the discovery of the axoneme (for review see~\cite{satir2014}).
Following the finding of ATPase distributed along the flagellum~\cite{nelson1958} and the conclusion from mechanical modelling that actuating elements must be distributed along its length~\cite{machin1958}, theoretical investigations initially focused on the possible role of contractile elements in generating motion. 
This work led to an early concept of \emph{curvature control}, i.e.\ that active moment generation is proportional to the local curvature~\cite{machin1963}. 
The contractile hypothesis was superseded by sliding (inextensible) filament theory~\cite{satir1965,satir1968}, which prompted the first computer simulation studies of flagellar movement~\cite{brokaw1971,brokaw1972}.
Although the contractile filament hypothesis has since been abandoned, in these early and subsequent studies, it was found that curvature control still provided an effective explanation for flagellar oscillation and wave propagation.
While early models~\cite{brokaw1971,brokaw1972} typically assumed that local moment is determined instantaneously by local curvature with a time delay (mathematically, of the form \(m(\kappa(t-\tau))\), where \(\kappa\) is curvature, \(t\) is time and \(\tau\) a time delay), later work formulated the process via the verbal description ``Oscillation results from a distributed control mechanism that reverses the direction of operation of the active sliding mechanism when the curvature reaches critical magnitudes in either direction''~\cite{brokaw1985}.
This description will form the basis for the mathematical modelling in the present paper, wherein we formulate a continuous dynamical system in the time domain, alongside recent advances in computational mechanics and biophysical modelling.

The phrase \emph{curvature control} has also been used in the literature for kinematically-actuated swimmer models that explicitly specify the waveform as opposed to modelling active moment generation and predicting the resulting waveform~\cite{alouges2013}.
Another popular and valuable approach to simulating flagellar movement is to specify a travelling wave of target curvature of a Kirchhoff rod, which has been used to investigate calcium dynamics, wall interactions~\cite{huang2018} and three dimensional flagellar waves~\cite{carichino2019}.
Furthermore, a number of authors have modelled components of the axoneme directly to investigate control mechanisms~\cite{dillon2000,dillon2003,dillon2007,bayly2016}.

Other models of flagellar regulation have been proposed and investigated computationally, for example Brokaw's formulation of piecewise linear dependence on rate of microtubule sliding~\cite{brokaw1975},  Hines and Blum's linear ordinary differential equation curvature control model~\cite{hines1978}, Lindemann's \emph{geometric clutch} model based on regulation by forces normal to the axoneme~\cite{lindemann1994,lindemann1994b,lindemann2002}, and Camalet and J\"{u}licher's frequency domain model with complex sliding resistance~\cite{camalet2000}.
In a comparative study of the modes of oscillation supported by each model, Bayly and Wilson found that a sliding controlled model would be dominated by retrograde modes, whereas a curvature control model based on that of Hines and Blum exhibited realistic anterograde waves~\cite{bayly2015}.
Their analysis was also supportive of a geometric clutch-type mechanism.
Sartori et al.~\cite{sartori2016} studied regulation of beating in model \emph{Chlamydomonas} flagella and found that regulation by the time derivative of curvature produced waveforms that most closely resembled observations.
More recently, the same group found that changes in curvature sensitivity can explain the modulation of \emph{Chlamydomonas} beat pattern shapes observed across changes in ATP, temperature, viscosity, microtubule stiffness, and several mutants~\cite{geyer2022}. 

These studies provide strong evidence for the relevance of curvature control as a generic model for axonemal bending regulation.
In this paper we focus on the specific case of the mammalian sperm flagellum.
Key areas of methodological development will include the use of a large amplitude, geometrically nonlinear model, with nonlocal hydrodynamics, analysed through direct simulation in the time domain, fitted to imaging data of human sperm.
Specific features of the mammalian sperm to take into account include the dramatic tapering of elastic stiffness from proximal to distal, and the inactive --- but mechanically critical --- end piece region~\cite{omoto1982,neal2020}.
This analysis will reveal how, despite the apparent internal complexity of the axoneme and its regulation, waveform changes across viscosity~\cite{smith2009} and activation state~\cite{ooi2014} to achieve efficient motility~\cite{ishimoto2018} can emerge from the the nonlinear dynamics of a parsimonious curvature control model.
In the language of nonlinear dynamics, we will see waveform switching behaviour as a result of a \textit{cusp catastrophe} --- the emergence of multiple co-existing solutions (flagellar waveforms in the present context) from a single solution as the underlying control parameters are varied.

\section{A simple mathematical model of active moment control}
            
    We consider the restricted planar motion of a human sperm flagellum, which provides a good approximation to motility modes observed in either spatially-restricted or high viscosity environments~\cite{smith2009,ooi2014}.
    Beating is driven by the internal sliding action produced by dynein activity, which can be coarse-grained as an internal \textit{active} moment per unit length, varying both in time and along the length of the flagellum~\cite{brokaw1971}. 

    \begin{figure}[tp]
       \centering
       \includegraphics[width=0.5\columnwidth]{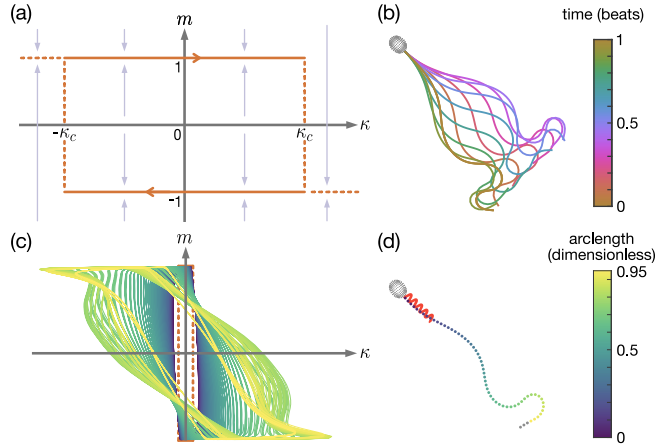}             
       \caption{Overview of the ARCH model, showing: (a) a sketch of the active moment density \(m\) against curvature \(\kappa\), for critical curvature \(\kappa_c\) (arrows show  direction of \(\dot{m}\)~\eqref{eqn:moment}); and (b)--(d) results from a single simulation with parameters \((\mathcal{S},\rho,\mathcal{M},\kappa_c) = (18, 36.4, 900, 1.5)\), comprising: (b) the flagellar waveform over a single beat; (c) active moment against flagellar curvature over 5 beats; (d) swimming track and flagellar position after 5 beats. Panels (c) and (d) share the same colour-scheme, depicting the change in active moment along the non-dimensionalised arclength of the flagellum.}
       \label{fig:progressive}
    \end{figure}

    In the physiological cell, the timescale associated with molecular motor attachment entails that a time delay must exist between the curvature threshold being reached and the switching of neighbouring components on the axoneme~\cite{lindemann2010flagellar}.
    Accounting for this hysteretic relationship, below we develop a simple mathematical model of active moment control, wherein a switch in sign of the rate of change of active moment density occurs on reaching a threshold curvature.

       Assuming that a fully-engaged section of dynein activity along the axoneme produces a preferred moment density \(m_c\), we construct a rate equation for the active moment density along the length of the flagellum, hereafter termed the \textit{Axonemal Regulation of Curvature, Hysteretic} model (ARCH). 
       After rescaling model equations with respect to flagellum length for position, inverse switching rate for time and \(m_c\) for force, the model takes the dimensionless form,
       \begin{align}
              \dot{m}\left(s, t\right) &= 
              - \left(m - 2\, \text{sgn}\left(m\right)\right)
              \mathcal{H}\left(\kappa_c - \text{sgn}\left(m\right)\kappa\left(s,t\right)\right)\nonumber\\
              &\qquad - \text{sgn}\left(m\right),
              \label{eqn:moment}
       \end{align}
       for the active moment density \(m\left(s,t\right)\) and curvature \(\kappa\left(s,t\right)\), as functions of curvilinear arclength \(s\) and time \(t\); the parameter \(\kappa_c\) denotes dimensionless critical curvature threshold and the functions \(\text{sgn}\left(\cdot\right)\) and \(\mathcal{H}\left(\cdot\right)\) represent the signum and Heaviside functions respectively. For more details on the derivation see the Supplementary Material section~S1.
       A phase-plane of the the model is shown in Figure~\ref{fig:progressive}a.

       The ARCH model drives the emergence of elastic waves along the flagellum, which can be described (following~\cite{neal2020}) by a constitutively linear but geometrically nonlinear filament, with the addition of the hydrodynamic force density \(\mathbf{f}\) exerted by the filament onto the fluid.
       Following~\cite{gaffney2011}, the elastic stiffness of the filament decreases from the proximal end (at the junction between the head and flagellum) to the distal tip (Eqn.~\ref{eqn:stiffness}), with distal stiffness \(E_d = 2.2\times 10^{-21}\)~Nm\(^2\) and dimensionless proximal/distal stiffness ratio \(\rho\). 
       The active moment, described by~\eqref{eqn:moment}, acts along the entirety of the flagellum, with the exception of the inactive distal end piece (where the doublet-microtubule structure of the axoneme is lost, see~\cite{neal2020}), specified here to be \(5\%\) of the total flagellum length.

       Human sperm swim in a very-low Reynolds number fluid environment which, neglecting non-Newtonian effects, can be described by the Stokes flow equations (see Materials and Methods).
       The nonlocal hydrodynamic effects associated with flagellum-head and flagellum-flagellum interactions are solved for utilising the method of regularized stokeslets~\cite{cortez2001,cortez2005}, employing the efficient nearest-neighbour discretisation~\cite{smith2018,gallagher2018} to model the three dimensional sperm head. 

       The full system (see Material and Methods for details), comprising active moment control, elastohydrodynamic equations and closed by force- and moment-free conditions over the cell is fully parameterised by the dimensionless groupings: \(\mathcal{S}\) (ratio of viscous to elastic forces), \(\rho\) (ratio of proximal to distal stiffness), \(\mathcal{M}\) (ratio of active to elastic forces) and \(\kappa_c\).
       For each choice of parameters \(\left(\mathcal{S}, \rho, \mathcal{M}, \kappa_c\right)\), the ordinary differential equation system~\eqref{eqn:ode_system} is solved using the MATLAB {\texttt{ode15s}} scheme until a stable beat pattern emerges.
       This approach gives rise to a progressive, human sperm-like, waveform (Fig.~\ref{fig:progressive}b) through the emergence of limit cycles in the active curvature density, varying along the flagellum (Fig.~\ref{fig:progressive}c-\ref{fig:progressive}d).

       \begin{figure*}[tb]
              \centering
              \includegraphics[width=\textwidth]{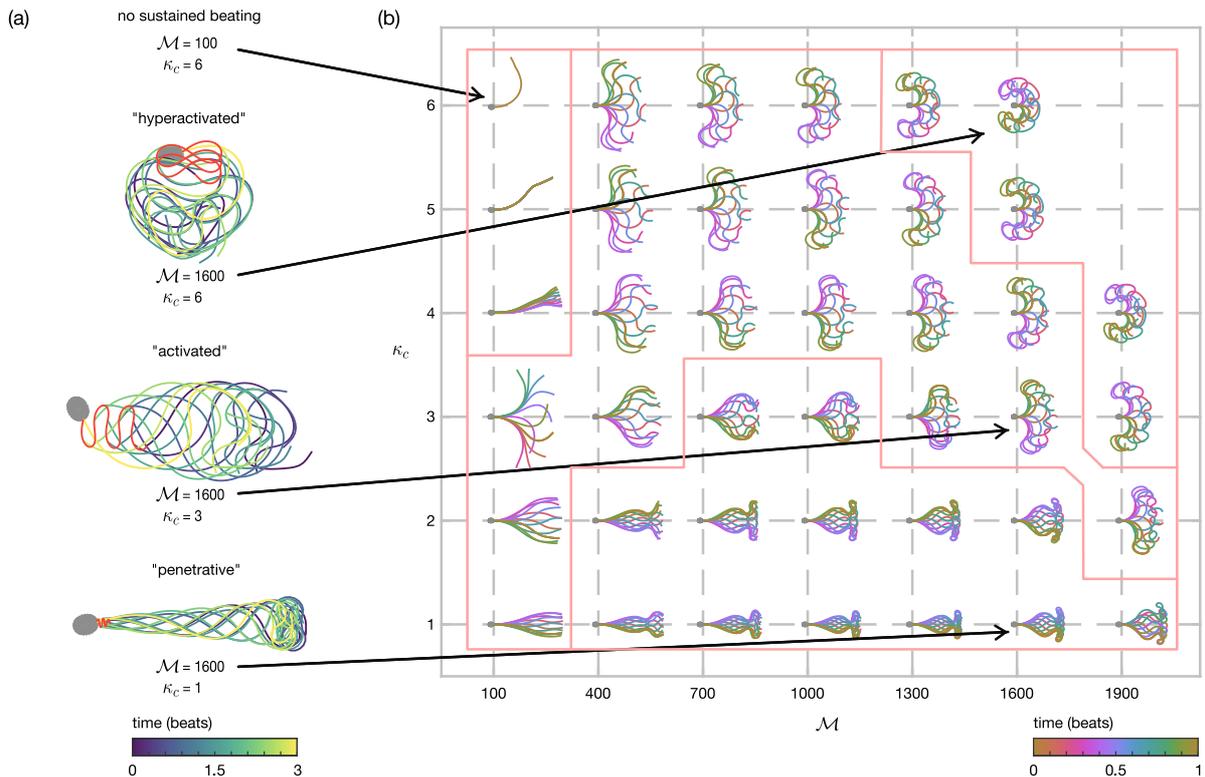}
              \caption{Sperm waveforms from simulations. (a) Three `characteristic modes' of flagellar beating and simulations with no sustained beating. (b) The active moment \((\mathcal{M})\) and critical curvature \((\kappa_c)\) parameters are varied with \((\mathcal{S}, \rho) = (18, 36.4)\). Parameter choices with missing entries (e.g. \(\left(\mathcal{M}, \kappa_c\right) = \left(100,5\right)\)) are unable to generate a beat.}
              \label{fig:waveform}
       \end{figure*}

\section{Simulations reveal variety of waveforms}

       Human sperm exhibit a variety of waveforms, notably depending on the surrounding fluid environment and biochemical signalling within the cell.
       This variety is often represented in terms of several \textit{characteristic modes} of motility, which can be categorised as: 
       1. \textit{penetrative} --- a low amplitude waveform with high curvature towards the end of the flagellum producing a track with little side-to-side yaw, often seen in sperm swimming in high-viscosity or shear-thinning fluids, important for penetrating cervical mucus; 
       2. \textit{activated} --- a moderate amplitude waveform producing a progressive track with significant yaw, often seen in sperm swimming in low-viscosity fluids; 
       and 3. \textit{hyperactivated} --- exhibiting greatly increased curvature compared to either the penetrative or activated waveforms~\cite{yanagimachi1970movement}, seen in sperm undergoing an influx of intracellular calcium, important for detachment from epithelial binding~\cite{alasmari2013ca2}.
       Examples of these characteristic modes are presented in Fig.~\ref{fig:waveform}a.

       To assess the diversity of waveforms that can arise from the ARCH model, computational simulations are performed over a range of parameter space, \(\mathcal{S} \in \left[12, 18\right]\), \(\rho \in \left[18.2, 36.4\right]\), \(\mathcal{M}\in\left[100, 2000\right]\), and \(\kappa_c\in[1, 10]\). 
       Results for the choice \(\left(\mathcal{S}, \rho\right) = \left(18, 36.4\right)\), with \(\left(\mathcal{M},\kappa_c\right)\) varied, are shown in Fig.~\ref{fig:waveform}b, with additional results in the Supplementary Information. 
       Crucially, we see that this simple model can exhibit the variety of \emph{planar} waveforms that are seen physiologically.
       For each choice of \(\mathcal{S}, \rho, \mathcal{M}\), small values of \(\kappa_c\) produce penetrative waveforms, with a step change to an activated waveform as \(\kappa_c\) is increased. Relatively large values of \(\kappa_c\) lead to the simulated sperm no longer swimming progressively, instead forming the large amplitude, increased curvature waveforms that are typical of hyperactivated sperm. 

       The full dataset (see Supplementary Information Figure~S4) shows the effect of each parameter on the simulated waveforms. 
       The proximal-distal stiffness ratio \(\rho\) exhibits an inverse relationship with the wavenumber of the emergent beat. The combination of swimming parameter \(\mathcal{S}\) and active moment parameter \(\mathcal{M}\) limit the region in parameter space in which progressive beats can form. 
       In particular, larger choices of \(\mathcal{S}\) (i.e.\ increased viscosity fluids \emph{at fixed flagellar stiffness and length}) sustain a greater region of penetrative waveforms. 
       This observation agrees with the experimental finding that the penetrative motility mode is more commonly seen in high-viscosity or shear-thinning fluids, which more closely represent the physiological environment of female cervical mucus.

       \begin{figure*}[ht]
              \centering
              \includegraphics[width=\textwidth]{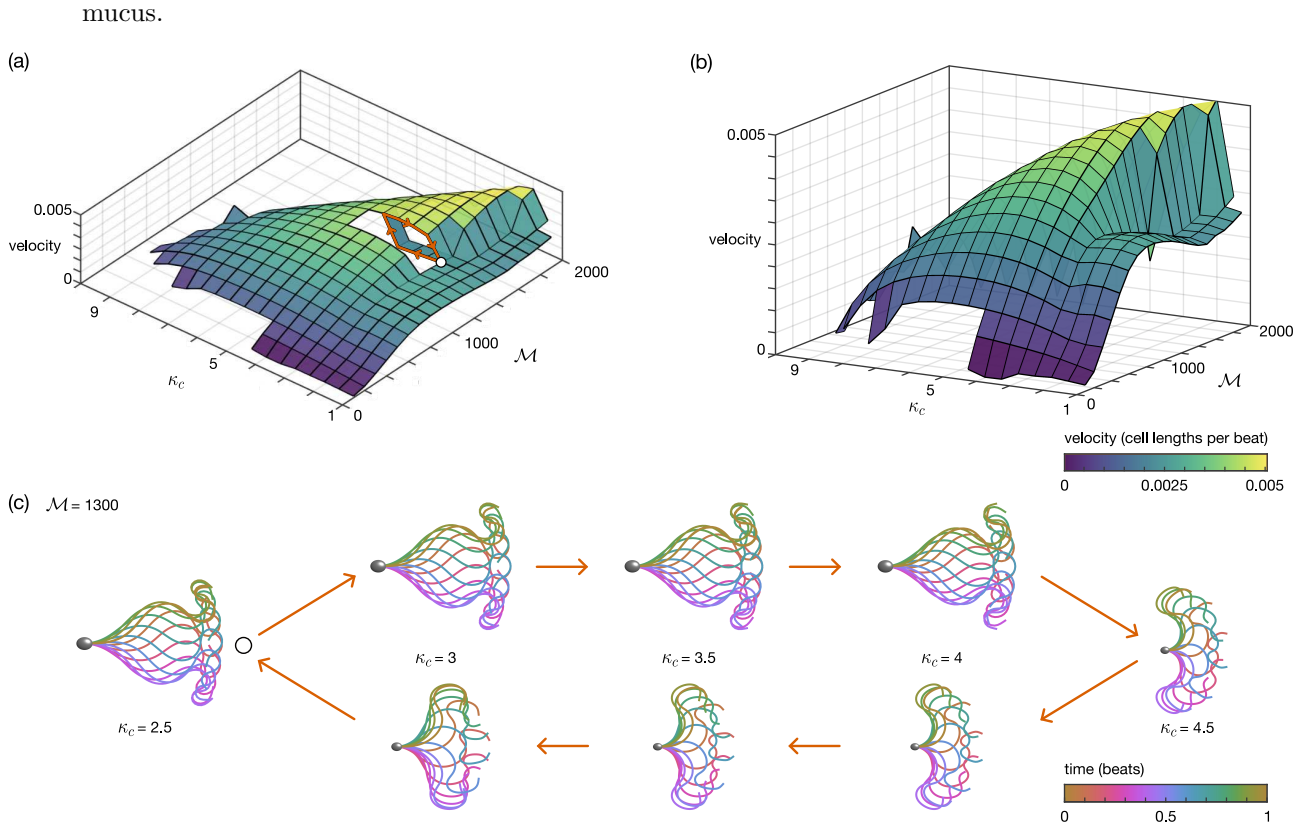}
              \caption{A cusp catastrophe bifurcation enables waveform switching. Panels (a) and (b) show the velocity (in cell lengths per beat) as the active moment \((\mathcal{M})\) and critical curvature \((\kappa_c)\) parameters are varied with \((\mathcal{S}, \rho) = (18, 36.4)\). In panel (a), a section of the surface is removed so that the lower branch of the bifurcation can be seen, with a path around the bifurcation shown in orange. In panel (b) the figure is rotated to highlight the cusp catastrophe. Finally, panel (c) details the hysteresis along the orange path in (a).}
              \label{fig:vel_ic}
       \end{figure*}
       
       The full results of the ARCH model in terms of the simulated sperm tracks (shown over 5 beats); sperm waveforms; sperm flagellar timelapse; dimensionless swimming velocity; and the dimensionless beat frequency are shown in the Supplementary Material, Figures~S3--S7.
       These simulations show that the ARCH model is able to recreate many of the physiological features of swimming human sperm, including non-progressive (yet still beating) waveforms, and simulations where waveforms do not develop.
       This finding is supportive of the concept of the curvature control hypothesis more generally, and provides additional evidence that curvature may play a significant role in the oscillatory switching of the axoneme. 
       The computationally lightweight nature of this model allows for detailed investigations into the underlying mechanisms, particularly, as we will see, relating to the switching of beating modes and comparisons with experimentally tracked cells.

       \section{Switching between penetrative and activated waveforms occurs via a cusp catastrophe bifurcation}

       An aspect of flagellar dynamics that is of particular interest is in the qualitative switching of behaviour between penetrative and activated waveforms.
       It has long been thought that the penetrative motility mode is characteristic of swimming in high-viscosity fluids~\cite{suarez1992}, although more recent studies suggest that shear-thinning and viscoelastic effects may play an important role~\cite{hyakutake2019}.
       In the results presented above (see Fig.~\ref{fig:waveform}), we have seen that, as \(\kappa_c\) is increased, there is a step-change in behaviour as the beat pattern switches between penetrative and activated types, which warrants further investigation.

       Fig.~\ref{fig:vel_ic} plots sperm velocity (in dimensionless units of cell lengths per beat) as \(\kappa_c\) and \(\mathcal{M}\) are varied for the choice \(\left(\mathcal{S},\rho\right) = \left(18, 36.4\right)\). 
       Small values of active moment parameter \(\mathcal{M}\) produce little progressive velocity due to the lack of development of a time-irreversible waveform (see Fig.~\ref{fig:waveform}b). 
       As \(\mathcal{M}\) is increased, a cusp catastrophe bifurcation emerges~\cite{thom1974stabilite}, heralding the formation of the time-reversible flagellar beats required for progressive swimming.
       For these values of \(\mathcal{M}\), as \(\kappa_c\) is increased from small to large, we can clearly observe the sharp increase in progressive velocity accompanying the qualitative change between penetrative and activated waveforms, before a smooth decrease to the non-progressive, hyperactivated, motilities.

       To characterise the behaviour in this region, we fix the choice of \(\mathcal{S}, \rho, \mathcal{M}\) and increment \(\kappa_c\) in the range \(\kappa_c\in[2.5,4.5]\), taking the initial shape at each step to be that of the previous simulation (as opposed to the parabolic initial condition used for all other simulations, see Materials and Methods).
       Following this scheme, first increasing \(\kappa_c\) in increments of \(0.5\) before iterating backwards, we see the emergence of hysteresis (shown in terms of velocity in the cut-out section of Fig.~\ref{fig:vel_ic}a, and in waveforms in Fig.~\ref{fig:vel_ic}c).
       It is this hysteresis in waveform for choices of critical curvature parameter \(\kappa_c\) that enables the switching behaviour seen in human sperm between penetrative and activated waveforms.

\section{Validation reveals a good fit to experimental data}

    Having shown the ability of the ARCH model to explain qualitative mode switching, we now address quantitative fitting to experimental data.
    To do this we have taken a tracked experimental cell corresponding to each of the three characteristic motility modes, found the choice of parameters that minimizes the sum squared difference in mean absolute curvature along the flagellum between simulation and experiment (for full details see Materials and Methods), and plotted the comparison waveforms in Fig.~\ref{fig:exp}.
    The qualitative features between the experimental and simulated cells are well replicated, with good agreement in the mean absolute curvature ranging between \(5\%\) and \(18\%\) error.

    \section{Discussion}

    ARCH provides a simple, yet physiologically representative model of the mechanism of flagellar regulation for controlling the formation and propagation of oscillations.
    By augmenting the non-local elastohydrodynamic partial differential equation system with a hysteretic rate equation for the evolution of the active moment density along the flagellum, we have shown that the active moment control model can produce the characteristic variety in waveforms that are seen in human sperm (Fig.~\ref{fig:waveform}). 
    Taking inspiration from the work of Brokaw~\cite{brokaw1971,brokaw1972,brokaw1985}, the key feature of ARCH is regulation of moment switching due to the local flagellar curvature, in contrast to approaches which utilise time delay~\cite{brokaw1971,brokaw1972}, prescribe the waveform~\cite{phan1987}, the shape of the moment activation function~\cite{lowe2003} or even employ a `target curvature'~\cite{fauci1995}. 
    In the present work, the archetypal sperm-like waveforms emerge entirely from the balance between the coupled elastic, hydrodynamic and internal force and moments rather than through specific choices made in the model. 

    \begin{figure}[t]
           \centering
           \includegraphics[width=0.5\columnwidth]{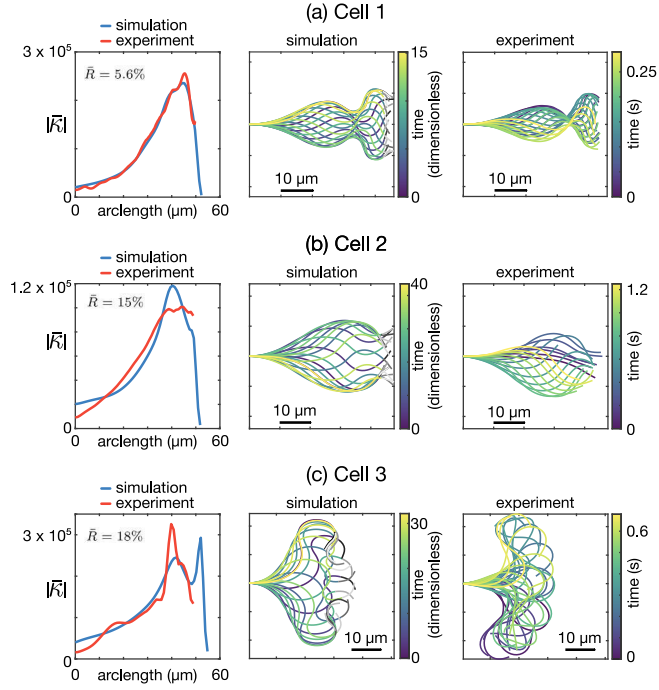}
           \caption{Comparison between simulation and experimental data for (a) penetrative, (b) activated, and (c) hyperactivated waveform types. The first column plots the mean absolute curvature along the flagellum for both simulation (blue) and experiment (red) against arclength in \(\mu m\). The second and third columns show the simulated and experimental waveforms respectively. In each panel, the error measure is calculated via \(\bar{R} = {({\sum(\lvert\bar{\kappa}_{\text{sim}}\rvert - \lvert\bar{\kappa}_{\text{exp}}\rvert)}^2)}^{1/2} / {(\sum \lvert\bar{\kappa}_{\text{sim}}\rvert^2)}^{1/2}\), where \(\bar{\kappa}\) is averaged over a single beat and the summation is along the arclength of tracked flagellum.}
           \label{fig:exp}
    \end{figure}

    Simulations over a range of dimensionless parameters \((\mathcal{S},\rho,\mathcal{M},\kappa_c)\) revealed the emergence of a cusp catastrophe bifurcation which, together with the associated hysteresis, enables flagellar waveform switching between penetrative and activated motility modes.
    This property of the active control model may explain how human sperm are able to have qualitatively different patterns of flagellar motion, while having the same regulatory parameters, thereby without the need for external signalling cues.
    Furthermore, changing the fluid environment (in this case by varying the viscous/elastic ration \(\mathcal{S}\)), while keeping regulatory parameters \((\mathcal{M},\kappa_c)\) fixed, can switch motility modes, mirroring what has been widely reported experimentally~\cite{smith2009,gallagher2019rapid} (results in Supplementary Material, Fig.~S8).
    In contrast, the hyperactivated motility mode emerges from model simulations in which the critical curvature threshold \(\kappa_c\) is greatly increased relative to the penetrative and activated motilities.
    This behaviour of the model is consistent with our interpretation of previously-published experimental data, as follows: hyperactivation occurs physiologically when progesterone induces calcium uptake into the sperm through the Catsper channel~\cite{sun2017catsper} leading to an increase in ATP production~\cite{suarez2008control}. 
    Flagellar bending occurs through the walking (detachment and reattachment) of dynein motors between neighbouring pairs of microtubule doublets in the axoneme, requiring ATP hydrolysis~\cite{lindemann2021many}.
    The increase in ATP production occurring during hyperactivation, may therefore be able to sustain greater bending (an increase in critical curvature threshold \(\kappa_c\)) than in the unstimulated progressive/activated states.

    It is noteworthy that the ARCH model is capable of producing simulations of both motile progressive (activated, penetrative) and motile non-progressive (hyperactivated) sperm, as well as regions where no sustained beating arises (see Fig.~\ref{fig:waveform}, and Supplementary Material Figs.~S3--S7). 
    These findings show that it is a balance between the active moment and critical curvature threshold which drives swimming. 
    In the case that the active moment is too small (\(\mathcal{M} = 100\), \(\kappa_c > 3\)) an equilibrium is found where no flagellar beat arises. 
    Decreasing \(\kappa_c\) then gives rise to a Hopf-bifurcation and the emergence of periodic beating. 
    The balance between these parameters also governs both the location of the cusp catastrophe (which characterises switching between the active and penetrative motilities), and the onset of the hyperactivated state. 
    The region of large critical curvature (\(\kappa_c\)) relative to active moment (\(\mathcal{M}\)) (e.g. \(\mathcal{M} = 1900\), \(\kappa_c = 5\), Fig.~\ref{fig:waveform}) lies beyond the scope of the present formulation, wherein the flagellum wraps around the sperm head causing self-intersection (see Supplementary Material Fig.~S9).
        
    The results presented in this manuscript show good agreement to flagellar waves observed in imaging experiments (Fig.~\ref{fig:exp}), demonstrating the potential of this model in extracting mechanistic insight from observed in experiments involving physiological and pharmacological modulation. 
    The nature of the active moment control model should allow for simple modifications to further capture the physiological realities of sperm motion, which we discuss below.
    
    Waveforms observed experimentally in human sperm tend not to be perfectly symmetrical in the cell-frame, but to have some inherent asymmetry due to the internal structure of the axoneme.
    A natural extension, therefore, is to modify the active moment control model to have disparate critical thresholds depending on the sign of the curvature.
    What effect the degree of asymmetry has on the resulting sperm trajectories would be of interest; in particular can the combination of asymmetry and the presence of boundaries combine to replicate the circular swimming paths that are often seen experimentally?
    
    \sloppy{Although planar flagellar waves are commonly observed, particularly in physiologically-representative high viscosity and/or spatially confined environments~\cite{nosrati2015two,gallagher2018casa}, a further area of development for the model would be to address the three-dimensional model seen by sperm in bulk fluid at low viscosity~\cite{su2012high,su2013sperm}.}
    In the first instance, adaptation of the elastohydrodynamic methodology to account for elastic twisting may allow for simulations of the planar-bending-with-twist motion encountered, for example, in hamster spermatozoa~\cite{woolley1977evidence}.
    
    In the present manuscript we have focussed on the application of the active moment control model to human-like spermatozoa.
    However, the internal `9 + 2' axonemal structure of the flagellum of human sperm is largely ubiquitous across species, with the main difference being the differing external accessory structures which can be modelled using modifications to the elastic structure of the simulated filament. 
    Furthermore, it is known that head morphology may play a significant role in shaping the flagellar waveform.
    As such, there may be significant cross-species interest in understanding how whether this control mechanism is able to recreate what is seen across a wide-range of wildlife, including other mammalian species, fish and birds~\cite{van2020computer,yanagimachi2022mysteries}.

    \section{Materials and Methods}

    \section{Elastohydrodynamic modelling of a human sperm}
    
        We model a sperm as comprising a rigid body (head) coupled to a planar elastic filament (flagellum). 
        We approximate the head of a human spermatozoa as an ellipsoidal surface (\(\partial H\)) with axes of length \(0.04\), \(0.032\), and \(0.02\), joined at the point \(\mathbf{X}_0(t)\) with angle \(\phi(t)\) to a planar flagellum of unit length. 
        Parametrising the flagellum by arclength \(s\) (with \(s = 0\) corresponding to the head/flagellum joint, and \(s=1\) to the distal tip), the position of the flagellum \(\mathbf{X}(s,t)\) can be fully described in terms of the angle between the tangent and the laboratory frame horizontal \(\theta\left(s,t\right)\), 
        
        \begin{equation*}
               \mathbf{X}(s,t) = \mathbf{X}_0(t) + \int\limits_0^s{\left[\cos{\theta(s^{\prime},t)}, \sin{\theta(s^{\prime},t)}, 0\right]}^T \mathrm{d}s^\prime.
        \end{equation*}

        Following the method employed in previous work~\cite{hall2019,neal2020}, application of force and moment free boundary conditions at \(s=1\) yields the dimensionless elastohydrodynamic integral equation relating the shape of the flagellum to the internal and hydrodynamic moments, namely 
        
        \begin{align}
               &E(s)\partial_s\theta(s,t) - \mathcal{M}\int\limits_{s}^{1} m(s^{\prime},t)\mathcal{H}\left(\ell - s^\prime\right)\mathrm{d}s^{\prime}\nonumber\\
               &\qquad+ \mathcal{S}^4\mathbf{e}_3\cdot\int\limits_{s}^{1}\left[\mathbf{X}(s^\prime,t) - \mathbf{X}(s,t)\right]\times\mathbf{f}(s^{\prime},t)\mathrm{d}s^{\prime} = 0,
               \label{eqn:elastic}
        \end{align}
        for the active moment per unit length \(m\) and the hydrodynamic force per unit length exerted by the flagellum on the surrounding fluid \(\mathbf{f}\), with the dimensionless elastic stiffness, which varies along the length of the flagellum, described by
        
        \begin{equation}
               E(s) = \begin{cases}
                             \left(\rho - 1\right){\left(\frac{s - d}{d}\right)}^2 + 1, & s \leq d,\\
                             1, & d < s \leq 1.
                      \end{cases}
               \label{eqn:stiffness}
        \end{equation}
        The dimensionless parameters describe: the ratio of active to elastic forces (\(\mathcal{M} = L^2 m_c / E_d\)), the ratio of viscous and elastic forces (\({S = L{(\mu/E_d\tau_d)}^{1/4}}\)), the proximal-distal stiffness ratio (\(\rho = E_p/E_d\)), and the proportion of the flagellum over which the stiffness varies, \(d\), each written in terms of the dimensional scales: the length of the flagellum, \(L\); the preferred moment density, \(m_c\); the fluid dynamic viscosity, \(\mu\); the characteristic bending-switching time, \(\tau_d\); and the flagellar stiffness at proximal and distal ends, \(E_p\) and \(E_d\) respectively.
        Finally, the proportion of the flagellum that is actively bending is represented by the physiological constant \(\ell = 0.95\).
        
        At very low Reynolds number, the behaviour of the Newtonian fluid environment is governed by the Stokes flow equations
        \begin{equation*}
               -\boldsymbol{\nabla} p + \mu \nabla^2\mathbf{u} = \mathbf{0},\qquad\boldsymbol{\nabla}\cdot\mathbf{u} = 0,
        \end{equation*} 
        where \(p = p(\mathbf{x},t)\) is pressure, \(\mathbf{u} = \mathbf{u}(\mathbf{x},t)\) is fluid velocity and \(\mu\) is dynamic viscosity.
        The Stokes flow equations are coupled with the no-slip, no-penetration boundary condition over the filament, \(\mathbf{u}(\mathbf{X}(s,t),t) = \partial_{t}\mathbf{X}(s,t)\).
        In solving this problem numerically, we apply the `NEAREST' approach~\cite{smith2018,gallagher2018} for the method of regularised stokeslets~\cite{cortez2001,cortez2005}, wherein we instead consider the exactly incompressible Stokes flow equations driven by a spatially-concentrated smoothed force, namely
        \begin{equation}
               -\boldsymbol{\nabla} p + \mu \nabla^2\mathbf{u}  + \psi_\varepsilon\left(\mathbf{x}-\mathbf{y}\right)\hat{\boldsymbol{e}}_j = \mathbf{0},
               \label{eqn:stokes}
        \end{equation}
        where \(\varepsilon\ll 1\) is a regularisation parameter, \(\mathbf{y}\) is the location of the force, \(\mathbf{x}\) is the evaluation point, \(\hat{\boldsymbol{e}}_j\) is the basis vector in the \(j\)-direction and \(\psi_\varepsilon(\mathbf{x})\) is a family of `blob' functions that approximate the three-dimensional Dirac delta distribution \(\delta(\mathbf{x})\) as \(\varepsilon\rightarrow 0\).
        In particular the choice
        \begin{equation*}
               \psi_\varepsilon(\mathbf{x}) = \frac{15\varepsilon^4}{{\left(\lvert\mathbf{x}\rvert^2 + \varepsilon^2\right)}^{7/2}},
        \end{equation*}
        yields the well-known Green's function for~\eqref{eqn:stokes}, the Regularised Stokeslet~\cite{cortez2005},
        \begin{equation*}
               S_{ij}^{\varepsilon}(\mathbf{x},\mathbf{y}) = \frac{1}{8\pi\mu}\left(\frac{\delta_{ij}(r^2 + 2\varepsilon^2) + r_i r_j}{r_\varepsilon^3}\right),
        \end{equation*}
        where \(r_i = x_i - y_i\), \(r^2 = r_i r_i\), \(r^2_\varepsilon = r^2 + \varepsilon^2\) and repeated indices are summed over \(\{1,2,3\}\).
        
        Following the method laid out in~\cite{neal2020}, we then write the hydrodynamic equations for the motion of the head as
        
        \begin{align}
               &\mathbf{e}_j\cdot\left[\dot{\mathbf{X}}_0(t) + \dot{\phi}(t)\mathbf{e}_3\times\left(\mathbf{Y}(t) - \mathbf{X}_0(t)\right)\right]\nonumber\\
               &\qquad= \iint\limits_{\partial H (t)} S_{jk}^{\varepsilon}\left(\mathbf{Y}, \mathbf{Y}^\prime\right)\varphi_k\left(\mathbf{Y}^{\prime},t\right)\mathrm{d}S_{\mathbf{Y}^\prime}\nonumber\\
               &\qquad\qquad+ \int\limits_0^1 S_{jk}^\varepsilon\left(\mathbf{Y},\mathbf{X}(s,t)\right) f_k(s,t)\mathrm{d}s,\quad\text{for}\quad \mathbf{Y}\in\partial H(t),
               \label{eqn:hydro-head}
        \end{align}
        for points \(\mathbf{Y}\) on the surface of the head \(\partial H(t)\), which exerts force per unit area \(\boldsymbol{\varphi}(\mathbf{Y},t)\) on the surrounding fluid; and for the motion of the flagellum as
        
        \begin{align}
               &\mathbf{e}_j\cdot \dot{\mathbf{X}}(s,t) = \iint\limits_{\partial H(t)} S_{jk}^{\varepsilon}\left(\mathbf{X}(s,t), \mathbf{Y}\right)\varphi_k\left(\mathbf{Y},t\right)\mathrm{d}S_{\mathbf{Y}}\nonumber\\
               &\qquad + \int\limits_0^1 S_{jk}^\varepsilon\left(\mathbf{X}(s,t),\mathbf{X}(s^\prime,t)\right) f_k(s^\prime,t)\mathrm{d}s^\prime,\quad\text{for}\quad s\in [0,1].
               \label{eqn:hydro-tail}
        \end{align}
        
        The full system of equations~\eqref{eqn:moment},~\eqref{eqn:elastic},~\eqref{eqn:hydro-head},~\eqref{eqn:hydro-tail}, supplemented with the force- and moment-free conditions
        \begin{align}
            \iint_{\partial H(t)} \mathbf{\varphi}\left(\mathbf{Y}\right){d}S_{\mathbf{Y}} + \int\limits_{0}^{1} \mathbf{f}\left(s,t\right){d}s &= 0,\label{eqn:ff}\\
            \mathbf{e}_3\cdot\left[\iint_{\partial H(t)} \left(\mathbf{Y} - \mathbf{X}(0,t)\right)\times\mathbf{\varphi}\left(\mathbf{Y}\right){d}S_{\mathbf{Y}}\right.&\nonumber\\
            \left. + \int\limits_{0}^{1} \left(\mathbf{X}(s,t) - \mathbf{X}(0,t)\right)\times\mathbf{f}\left(s,t\right){d}s\right] &= 0,\label{eqn:mf}
        \end{align}
        are numerically discretised following Refs.~\cite{hall2019,neal2020}, with the non-local hydrodynamics accounted for using the parallelised NEAREST method of Smith and Gallagher~\cite{smith2018, gallagher2018, gallagher2020,gallagher2020b, gallagher2021}, culminating in an ODE system of the form (see Supplementary Material for details),
        \begin{equation}
               A \dot{\mathbf{z}} = \mathbf{b},
               \label{eqn:ode_system}
        \end{equation}
        for the trajectory, head angle, discretised flagellum, active moment, and force distributions, which subsequently can be solved in {\tt{MATLAB}\textsuperscript{\textregistered}} using the built-in solver {\tt{ode15s}}.
        
        Unless specified otherwise, the simulations in this manuscript are initialised with a flagellar shape corresponding to unit-arclength section of a low-amplitude parabola, sampled from the curve \({y = 0.1 x^2}\).
        
        For some choices of parameter groupings \((\mathcal{S},\rho,\mathcal{M},\kappa_c)\), initialising cells in this way can lead to self-intersection of the flagellum. 
        To prevent this issue, we introduce a repulsive force, as used in Refs.~\cite{brady1985,brady1988,ishikawa2007}. Decomposing the force term into hydrodynamic and repulsive components, i.e. \(\mathbf{f} = \mathbf{f}^{\text{hyd}} + \mathbf{f}^{\text{rep}}\). For a point on the flagellum at arclength \(s\), we write
        
        \begin{align*}
               \mathbf{f}^{\text{rep}}(s,t) &= 
               \left(\int\limits_{0}^{s - \delta} + \int\limits_{s+\delta}^1\right)
               A\frac{\exp{(- B d\!\left(s, s^{\prime}\right))}}{\left(1 - \exp{(- B d\!\left(s, s^{\prime}\right))}\right)}
               \frac{\mathbf{d}(s, s^{\prime})}{d\!\left(s, s^{\prime}\right)}
               \mathrm{d}s^{\prime}\\
               & + \int_{\partial H} 
               A\frac{\exp{(- B D\!\left(s, \mathbf{Y}\right))}}{\left(1 - \exp{(- B D\!\left(s, \mathbf{Y}\right))}\right)}
               \frac{\mathbf{D}(s, \mathbf{Y})}{D\!\left(s, \mathbf{Y}\right)}
               \mathrm{d}\mathbf{Y},
        \end{align*}
        where \(d(s,s^{\prime}) = \lvert\mathbf{x}(s,t) - \mathbf{x}(s^{\prime},t)\rvert\) is the distance between points on the flagellum, \(D(s,\mathbf{Y}) = \lvert\mathbf{x}(s,t) - \mathbf{Y}(t)\rvert\) is the distance between a point on the flagellum and a point on the head, the exclusion region \(\{s-\delta,s+\delta\}\) is chosen to prevent repulsion of neighbouring segments on the flagellum, and the dimensionless strength parameters \({A = 100}\) and \({B = 4 / \varepsilon}\) chosen via numerical experiment. To improve numerical performance, all points separated by a distance greater than \(4\varepsilon\) are set to have no contribution to \(\mathbf{f}^{\text{rep}}\).

        \section{Experimental procedures}
        
        The characteristic sperm presented in Fig.~\ref{fig:exp} are taken from a semen sample, obtained through masturbation following 2--3 days' abstinence, provided by an unscreened normozoospermic donor recruited at Birmingham Women's and children's NHS Foundation Trust after giving informed consent. 
        The experimental procedures for obtaining each type of waveform (activated, progressive, hyperactivated) are as follows:
        \begin{enumerate}
               \item Activated waveform:
        \end{enumerate}
        Cells were selected using a swim-up, whereby a 500~\(\mu\)L aliquot of Gamete Buffer (K-SIGB-100 Sydney IVF) was placed in a 5~mL round-bottom tube (Corning, Falcon 352058). 
        A 200~\(\mu\)L aliquot of semen was pipetted to the bottom of the tube, inclined and left in the incubator for 30 minutes at \(37^\circ\)C in 6\% CO\textsubscript{2}, after which a 200~\(\mu\)L aliquot was taken from the top of the solution and the remainder discarded. Cells were imaged in a 10~\(\mu\)m depth chamber (10-01-04-B-CE;\@ Leja Products B.V., Nieuw-Vennep, The Netherlands).
        
        \begin{enumerate}
               \setcounter{enumi}{1}
               \item Progressive waveform:
        \end{enumerate}
        Cells were suspended in Gamete Buffer (K-SIGB-100 Sydney IVF) with the addition of 1\% methylcellulose (M0512, Sigma-Aldrich, Poole, UK) specified so that an aqueous 2\% solution gives a nominal viscosity of 4000 centipoise or 4Pa s at \(20^\circ\)C.
        The cells were loaded by capillary action into flat-sided borosilicate capillary tubes (VITROTUBES, 2540, Composite Metal Services, Ilkley, UK) with length 50~mm and inner dimensions 4\(\times\)0.4~mm.
        One end of the tube was sealed with CRISTASEAL (Hawksley, Sussex, UK \#01503-00).
        Cells were selected for imaging by immersing the open end of the capillary tube into a 1.5~mL BEEM capsule (Agar Scientific, UK) containing a 200~\(\mu\)L aliquot of raw semen.
        Incubation was performed for 30 minutes at \(37^\circ\)C in 6\% CO\textsubscript{2}.
        Cells were imaged at 2~cm migration distance into the capillary tube and in the surface accumulated layer 10--20~\(\mu\)m from the inner surface of the capillary tube.
        
        \begin{enumerate}
               \setcounter{enumi}{2}
               \item Hyperactivated waveform:
        \end{enumerate}
        Cells were selected using a swim-up, following the same procedure as for the activated waveform.
        An aliquot of progesterone P4 (Sigma P-0310) at 10~mM in DMSO was mixed into the swim-up at a dilution of 1:20 followed by 1:100 to stimulate hyperactivation.
        Cells were imaged in a 10~\(\mu\)m depth chamber (10-01-04-B-CE;\@ Leja Products B.V., Nieuw-Vennep, The Netherlands).

    \section{Imaging and flagellar tracking}

    All cells were imaged using a Nikon Eclipse 80i microscope and negative-phase contrast microscopy at 10x magnification (10x 0.2 Ph1 BM \(\infty\)/-WD7.0) and an Andor Zyla 5.5 (Andor, Oxford UK) CMOS camera with pixel size 6.5~\(\mu\)m at 200~Hz for 3 seconds.
    Microscopy videos were then loaded into the FAST software package (\href{www.flagellarcapture.com}{www.flagellarcapture.com})~\cite{gallagher2019rapid} to extract the flagellar waveforms.

    \section{Fitting procedure for comparing simulation and experiment}

    A successful fitting procedure must take into account the complexity of the time- and space-varying properties of the flagellar waveform; the time at which a given beat `starts', the length of visible flagellum in experimental data, and the timescale over which beating occurs each must be addressed.
    A further complicating factor lies in the characteristic timescale \(\tau_d\), which in the dimensionless system (equations~\eqref{eqn:moment},~\eqref{eqn:elastic},~\eqref{eqn:hydro-head},~\eqref{eqn:hydro-tail}) is contained within the swimming number \(\mathcal{S}\), and is not immediately measurable from experimental findings. 
    
    Taking into account these potential difficulties, for each cell we calculate the mean (over three complete beats) of the absolute value of the curvature as it varies in arclength \(s\) (see Fig.~\ref{fig:exp}). 
    This measurement provides a time-independent characterisation of the flagellar waveform, which varies only in one-dimension (arclength), thus reducing the computational complexity of the fitting task.
    
    There is a significant computational expense involved in fitting experimental cells to the 4-parameter model, as each simulation requires solving until a steady beat is formed and maintained.
    As our aim with the fitting procedure is to validate that the active moment control model is capable of producing waveforms that are similar to those seen experimentally, we are not concerned with obtaining the true optimal parameter set that fit each experimental sperm.
    Instead, we perform simulations over a fixed range of parameter space discussed earlier (\(\mathcal{S} \in \left[12, 18\right]\), \(\rho \in \left[18.2, 36.4\right]\), \(\mathcal{M}\in\left[100, 2000\right]\), and \(\kappa_c\in[1, 10]\), as plotted in the data linked within the Supplementary Material) and select as a `fit' the choice of parameters (\(\mathcal{S}, \rho, \mathcal{M}, \kappa_c\)) that minimises (within the simulation dataset) the sum squared difference between the experimental and simulated calculations of mean absolute curvature.
    
    The experimental imaging and extraction of the flagellar waveform is not perfect; the distal end of the flagellum in particular is difficult to image and often moves out of plane when sperm swim. 
    As such, and to account for variations in flagellar length, we scale the simulated arclength by a length parameter \(l\in\left[30~\mu\text{m}, 60~\mu\text{m}\right]\) and subsample up to the length of the experimental flagellum for comparison.

\section{Acknowledgments}
A significant portion of the computational work described in this paper was performed using the University of Birmingham's BlueBEAR HPC, which provides a High Performance Computing service to the University's research community. See \href{http://www.birmingham.ac.uk/bear}{http://www.birmingham.ac.uk/bear} for more details.

\section{Supplementary Material}
Supplementary material is available at PNAS Nexus online.

\section{Funding}
The authors gratefully acknowledge funding from the Engineering and Physical Sciences Research Council (EPSRC) Healthcare Technologies Award (grant no. EP/N021096/1).
MTG acknowledges further financial support from the EPSRC via grant EP/N014391/2, the University of Birmingham Dynamic Investment Fund.
JKB and MTG were supported by the Medical Research Council [grant number MC\_PC\_19029].
JKB is funded by a National Institute of Health Research (NIHR), and Health Education England, Senior Clinical Lectureship Grant: The role of the human sperm in healthy live birth (NIHRDH-HCS SCL-2014-05-001).
This article presents independent research funded in part by the National Institute for Health Research NIHR and Health Education England.
The views expressed are those of the authors and not necessarily those of the NHS, the NIHR or the Department of Health.
The ongoing support of patients and staff at the Birmingham Women’s and Children’s NHS Trust is fundamental to our research work. 

\section{Author contributions statement}
M.T.G., J.C.K.B., D.J.S. designed the research and interpreted the results. M.T.G. implemented the computational model.

\section{Data availability}
The code for the methods set out in this manuscript are stored in GitLab: 
{\url{https://gitlab.com/nearest_code}} (containing individual repositories for the ARCH, NEARESTelastica and NEAREST packages). It should be noted that these repositories are under active development. The full data and code for this manuscript has been archived within the Zenodo repository: \url{https://doi.org/10.5281/zenodo.6866984}

  \bibliographystyle{plain}
  
\end{document}